# Identification of machining defects by Small Displacement Torsor and form parameterization method

Sergent A.[1], Bui-Minh H.[1], Favreliere H.[1], Duret D.[1], Samper S.[1], Villeneuve F.[2]

**(1)** : SYMME Laboratory, University of Savoie
5 chemin de Bellevue, 74944 Annecy le Vieux, France
Phone: +33450096572
Fax: +33450096543
E-mail : {alain.sergent, hien.bui-minh, hugues.favreliere, daniel.duret, serge.samper}@univ-savoie.fr

**(2)** : G-SCOP Laboratory
46 avenue Félix Viallet, 38031 Grenoble, France
Phone: +33476827031
Fax: +33476574695
E-mail : francois.villeneuve@g-scop.inpg.fr

**Abstract:** In the context of product quality, the methods that can be used to estimate machining defects and predict causes of these defects are one of the important factors of a manufacturing process. The two approaches that are presented in this article are used to determine the machining defects. The first approach uses the Small Displacement Torsor (SDT) concept [BM] to determine displacement dispersions (translations and rotations) of machined surfaces. The second one, which takes into account form errors of machined surface (i.e. twist, comber, undulation), uses a geometrical model based on the modal shape's properties, namely the form parameterization method [FS1]. A case study is then carried out to analyze the machining defects of a batch of machined parts.

## 1- Introduction and literature review

The machining defects are among the key factors that affect product quality. Nowadays, when the precision of machined parts is requested tighter, size and position deviations of products are considered along with orientation and form errors. The different approaches have been presented for analyzing geometrical defects in three-dimensional, such as in Kanai et al. [KO] where the matrix tool was used, while Clément and al [CL] used tensor modelling, and the Small Displacement Torsor (SDT) concept was developed by Bourdet [BM]. Thanks to the SDT concept, some authors developed simulation methods to predict manufacturing defects (positioning and machining defects) [AA, KF, KV, and TK]. However, the input data that was used to simulate these models was often theoretical data. As consequence, studies are able to provide experimental data are requested to verify the simulation models. An experimental approach was carried out by Legoff et al. [LV] in order to validate their 3D model on manufacturing tolerancing. Tichadou et al. [TL] presented a method that was used to determine the machining defects based on measured points of machined surfaces. There are some limitations in the above studies:

- The machining defects are obtained using some measured points on machined surfaces, which can't sufficiently represent defects of the analyzed surfaces.
- The methods that are used to reconstruct machined surfaces from measured points highly depend on the software of the coordinate measuring machine.
- The form defects have not been taken into account yet.

In the studies on form errors, Formosa et al. [FS2] proposed a method that was used to decompose form errors based on the dynamic natural modes, Samper et al. [SP] expressed that "As high precision assemblies cannot be analyzed with the assumption that form errors are negligible". Here, they analyzed a contact of a pair surfaces having form errors based on modal shape's properties.

In this paper, the two different approaches are used in order to identify machining defects of milled planes. The first one is used to determine translations and rotations of the machined planes based on the SDT concept. The second one takes into account form defects of the machined planes. The machining defects of a batch of 50 machined parts are analyzed using the two above approaches. Furthermore, some geometric characteristics of the machined planes can be obtained, i.e. flatness, parallelism. Finally, some comments are given on these results.

## 2- Small Displacement Torsor (SDT)

The Small Displacement Torsor (SDT) concept that was presented by Bourdet et al. [BM] is used to define geometrical defects of machined surfaces. The SDT concept is based upon the movements of a rigid body [EB]. Two vectors **R** and **T** are used to present three small rotations (rx, ry, rz) and three small translations (tx, ty, tz) and are gathered in a SDT. For instance, a SDT of a machined plane is used to express the relationships between the associated plane and the nominal plane in the origin of coordinates O (Figure 1). Let Z-axis be the normal of the nominal plane, the SDT of the machined plane therefore includes three components: two rotations around X and Y; one translation along Z. The torsor of this plane is shown in equation (1).

$$T_{Plane} = \{\mathbf{R} \quad \mathbf{T}\}_{(O,X,Y,Z)} = \begin{Bmatrix} rx & 0 \\ ry & 0 \\ 0 & tz \end{Bmatrix}_{(O,X,Y,Z)} \quad (1)$$

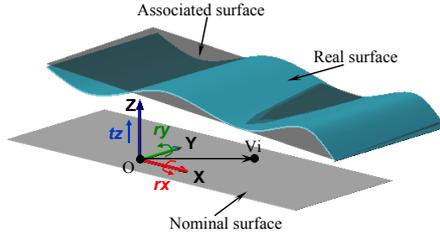

**Figure 1: SDT of a plane [KV]**

In this study, a batch of 50 machined parts is analyzed, consequently three SDT components of the machined planes will be variances of translation ($s_{T_Z}^2$) and rotations ($s_{R_X}^2, s_{R_Y}^2$).

## 3- Form parameterization method

The concept of this method is based on a discrete modal decomposition (DMD) [FS1]. The DMD decomposes a signal in a set of discrete functions, like a discrete Fourier transform. These signals are measured geometrical elements, for example: measured plane, measured cylinder, and measured sphere. The measured geometrical surfaces are searched and described in the set of discrete functions; some steps of this method can be shown as follows:
- The measured surfaces (measured points of the surfaces) are discretized using a finite element approach.
- These surfaces are expressed by a displacement vector (V)
- A modal analysis is used to obtain the modal basis ($Q_i$) that is then used to decompose the vector V and calculate the $\lambda_i$ coefficients, which represent the form deviation in the basis of modal shapes.
- Finally, the decomposition operation consists in projecting the vector V in the modal basis ($Q_i$).

$$Q^*.V = \left((Q^T.Q)^{-1}.Q^T\right).V = \lambda \quad (2)$$

where the basis of matrix Q is made of the vectors $Q_i$. The projection that is performed in the basis Q is not orthonormal, consequently the dual basis $Q^*$ has to be used.
It therefore come a new expression of vector V:

$$V = \sum_{i=1}^{m} \lambda_i.Q_i + \varepsilon \quad (3)$$

where m is the number of modes, which are chosen to present the measured vector V and the residual vector ε. The modes $Q_i$ are assessed through the resolution of a classic problem of mechanical vibration.

$$M.\ddot{q} + K.q = 0 \quad (4)$$

where
M is the matrix of generalized mass.
K is the matrix of generalized stiffness.
q is the displacement vector.

The resolution of this problem can be analyzed using the finite element method.
Thus, in the present method, the most significant modes of the modal basis are considered as representing the finite element of the machined planes. The obtained results can be used to evaluate some geometric characteristics of the machined plane, i.e. flatness and parallelism.

## 4- Experimental application

A batch of 50 work-pieces in aluminium that exhibit diameter of 30 mm and a length of 50 mm are machined using a CNC machine (DMG-Deckel Maho DMU 50). Measurements inside this machine are then carried out without disassembly the machined part out of the fixture. Two planes are machined by an end mill (ɸ20) with two different tool paths. Furthermore, the influence of the tool path on the machining defects is investigated. The first tool path that is used for machining the planes 1 is the spiral path. The other use for the planes 2 is a straight line path (Figure 2).

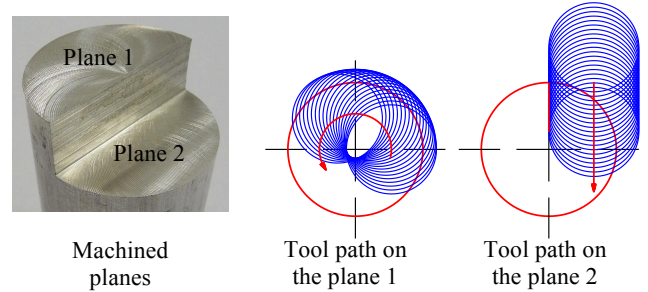

**Figure 2: Machined planes and tool paths**

Each machined plane is measured by the touch probe of this machine at 10 positions (measured points). The parameters of the measuring pattern are shown in figure 3.

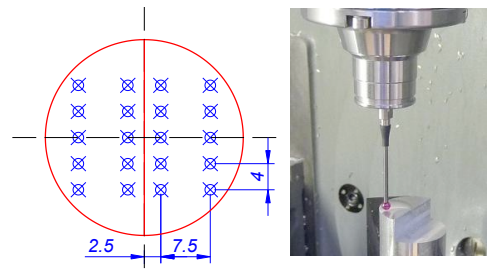

**Figure 3: the measuring pattern**

In order to ensure that noise measurements of this CNC machine do not influence the obtained results, two dimensions of a square gage block (class 0) were measured 800 times by Sergent et al. [SD]. Results show that the noise measurement on this machine is insignificant.

## 5- Presentation of results

### 5.1 – SDT method

Three components in a SDT of a machined plane (two rotations around X and Y; one translation along Z) will be calculated using the differences between an associated plane, which is a reconstructed plane, and a nominal plane (Figure 4). The best-fit least-square method, which is presented by Alistair [A], is used to reconstruct the machined planes from measured points.

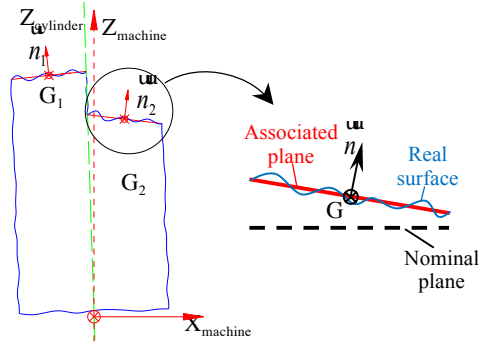

**Figure 4: Associated plane**

The SDTs components ($s^2_{R_X}, s^2_{R_Y}, s^2_{T_Z}$) of the machined planes are then obtained in order to estimate dispersions of machining. The results of the calculations are shown in tables 1 and 2, such as in figures 5 and 6.

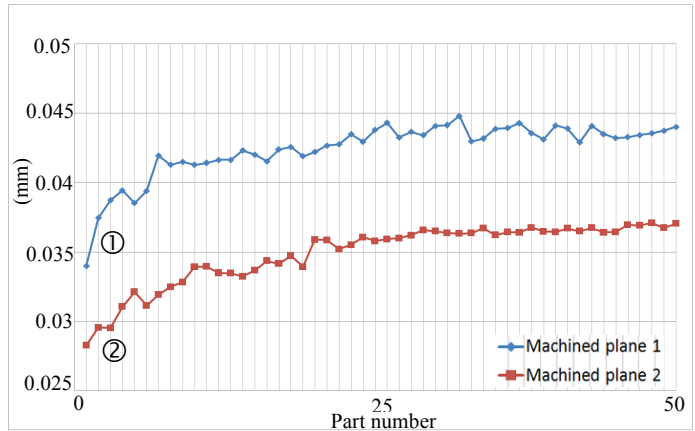

**Figure 5: Translations of the machined planes ($T_Z$)**

In figure 5, the curves number ① and number ② represent the translations along Z of the two milled planes. It can be seen that the translation defects of the two machined planes increase together during machining times (from the 1st part to the 50th part), namely drifts. The drifts may be due to temperature variations, which are not taken into account in the present study. These drifts can be called systematic errors and are corrected. The translations of the two machined planes after correction are shown in table 1.

| Name | Translations | |
|---|---|---|
| | $T_{Z1}$ | $T_{Z2}$ |
| Standard deviation (s)* (mm) | $7.406 \times 10^{-4}$ | $4.519 \times 10^{-4}$ |
| Variance ($s^2$)* ($mm^2$) | $5.486 \times 10^{-7}$ | $2.042 \times 10^{-7}$ |

**Table 1 : Translations of the machined planes**

*: values after corrections of systematic errors*

The differences of the rotations ($R_X$ and $R_Y$) of the planes are compared by the curves that have been put together in figure 6.

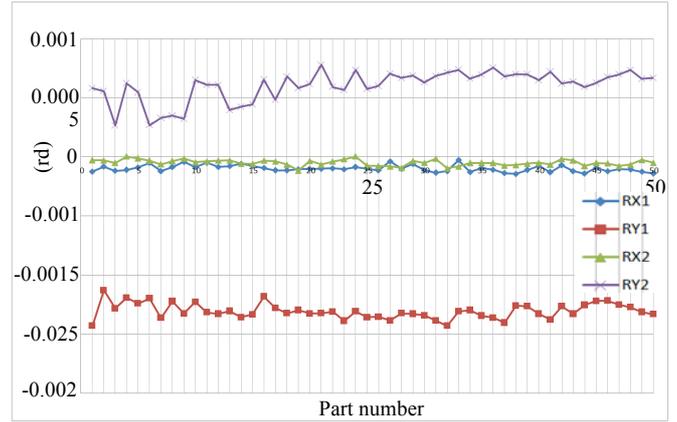

**Figure 6: Rotations of the machined planes**

| Name | Rotations | | | |
|---|---|---|---|---|
| | $R_{X1}$ | $R_{Y1}$ | $R_{X2}$ | $R_{Y2}$ |
| Standard Deviation (s) (rd) | $2.812 \times 10^{-5}$ | $6.459 \times 10^{-5}$ | $2.521 \times 10^{-5}$ | $12.795 \times 10^{-5}$ |
| Variance($s^2$) ($rd^2$) | $7.907 \times 10^{-10}$ | $41.7 \times 10^{-10}$ | $6.356 \times 10^{-10}$ | $163.7 \times 10^{-10}$ |

**Table 2 : Rotations of the machined planes**

The results show that the translation dispersion of the machined planes 1 ($s^2_{T_{Z1}} = 5.486 \times 10^{-7}$) is greater than the other one ($s^2_{T_{Z2}} = 2.042 \times 10^{-7}$). It can be explained by the spiral path on the machined planes 1, which is more complex than the straight line path on the machined planes 2.

### 5.2 – Form parameterization method

#### 5.2.1 – Determination of eigen-modes

This method focuses on form errors of the machined planes. Theoretical considerations, the residual vector ε equal zero if the number modes equals the number of measured points multiplied by the number of the degree of freedom. In order to reconstruct exactly the form errors of a surface, the number of modes ($N_{mode}$) is used to analyse different defects of the surface.

$$N_{mode} = n_p \times n_{df} \quad (5)$$

where

$N_{mode}$ is a number of calculated modes.
$n_p$ is a number of measured points.
$n_{df}$ is a number of degrees of freedom of each measured point.

In present study, only one degrees of freedom of the measured points (translation along Z) is considered. The finite element mesh is built up using the same number of nodes than the number of measured points (figure 7). The measuring pattern of this model has 10 nodes; therefore 10 modes can be obtained.

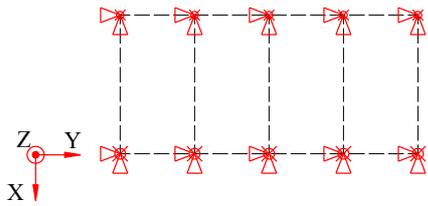

**Figure 7: Finite element model**

As previously mentioned, only the significant modes are considered for estimating the form defects. Thus, the eight significant modes are shown as in figure 8.

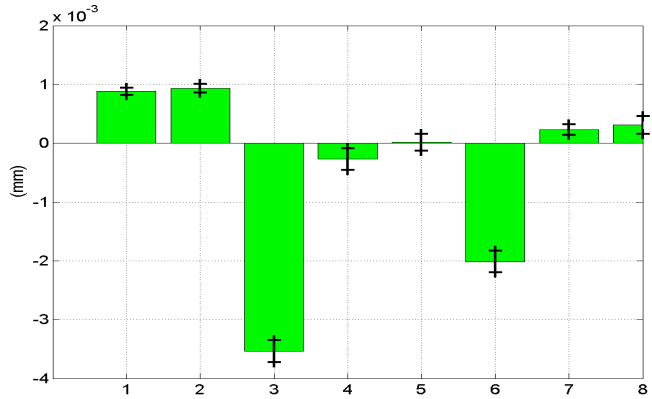

**Figure 8: Eight important modes of the machined planes 1**

They are named as follows.
- Mode 1 and 2:    translations and rotations
- Mode 3:          comber
- Mode 4:          1$^{st}$ twist
- Mode 5:          undulation
- Mode 6:          2$^{nd}$ twist
- Mode 7:          1.5 undulation
- Mode 8:          3$^{rd}$ twist

This method focus on the form errors of the machined planes, consequently the mode 3 (comber – figure 9a) and the mode 6 (2$^{nd}$ twist – figure 9b) of the machined planes 1 (figure 8) are the most significant modes of the form errors. Means and standard deviations of these two modes are calculated and shown in table 3.

| Name | Machined planes 1 | |
|---|---|---|
| | Mode 3 | Mode 6 |
| Mean (mm) | -3.535×10$^{-3}$ | -2.015×10$^{-3}$ |
| Standard deviation (mm) | 0.19×10$^{-3}$ | 0.185×10$^{-3}$ |

**Table 3: The most significant modes of the planes 1**

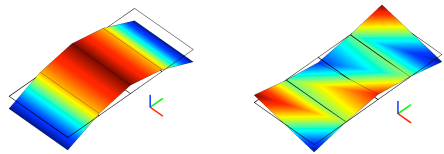

a) Mode 3         b) Mode 6

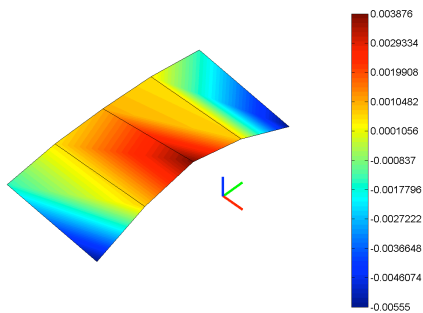

c) Mean of reconstructed forms

**Figure 9: Form errors of machined planes 1**

The two most significant modes are then used to obtain the final form errors (Figure 9c) of the machined planes 1. Similarly, the significant modes of the machined planes 2 are obtained and shown as in figure 10, and the means and standard deviations of the three most significant form modes are shown in table 4.

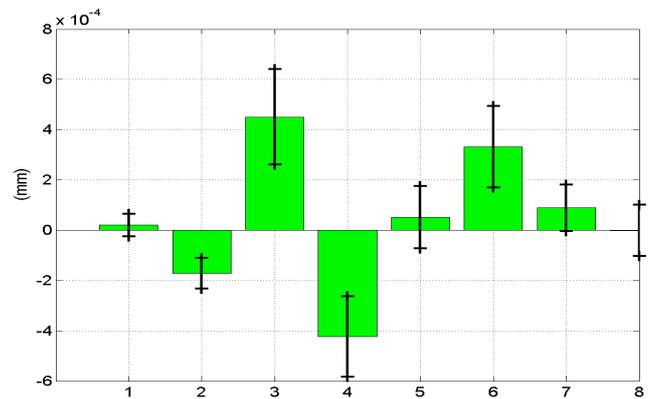

**Figure 10: Eight significant modes of the machined planes 2**

| Name | Machined planes 2 | | |
|---|---|---|---|
| | Mode 3 | Mode 4 | Mode 6 |
| Mean (mm) | 0.451×10$^{-3}$ | -0.423×10$^{-3}$ | 0.331×10$^{-3}$ |
| Standard deviation (mm) | 0.19×10$^{-3}$ | 0.161×10$^{-3}$ | 0.163×10$^{-3}$ |

**Table 4 : The most significant modes of the planes 2**

Figure 11d shows the form defects of the machined planes 2. It is obtained using the sum of the three most significant modes (3, 4 and 6 – Figures 11a, b, c).

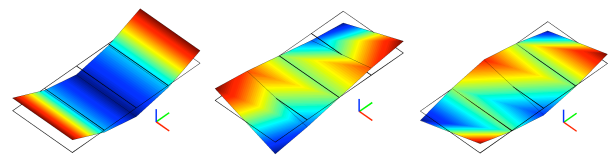

a) Mode 3         b) Mode 4         c) Mode 6

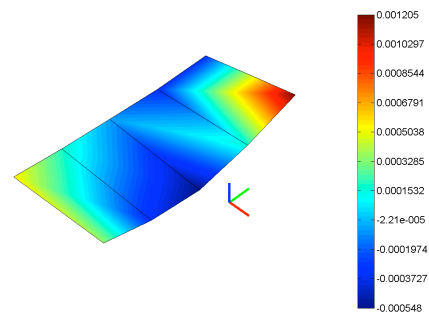

d) Mean of reconstructed forms

**Figure 11: Form errors of the machined planes 2**

Let us now compare the form errors of the two machined planes, it shows that:
- The form errors are greater on the planes 1 than on the planes 2.

- The form error of the planes 1 (Figure 9c) looks like a piece of a cone where the material that nears the centre is higher than the outside. It can be explained by the effect of the spiral path on these planes, this error did not appear on the planes 2 (Figure 11d).

### 5.2.2 – Determination of flatness and parallelism

Hereafter, the flatness and the parallelism of the machined planes can be used to verify the corresponding quality of these two planes.

| Name | Flatness | |
|---|---|---|
| | Planes 1 | Planes 2 |
| Mean $\bar{x}$ (mm) | $9.517\times10^{-3}$ | $1.943\times10^{-3}$ |
| Standard deviation $s$ (mm) | $0.458\times10^{-3}$ | $0.462\times10^{-3}$ |

**Table 5: Flatness values**

Table 5 shows the means and standard deviations of the flatness of the two machined planes. The distributions of the flatness and parallelism are shown in figure 12 and 13. The normality test states that the distributions of the flatness and parallelism are Gaussian.

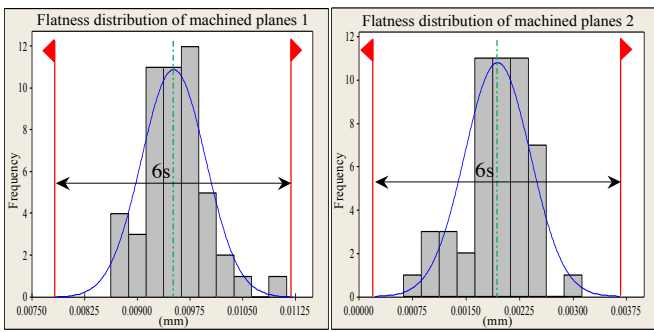

**Figure 12: Flatness distributions**

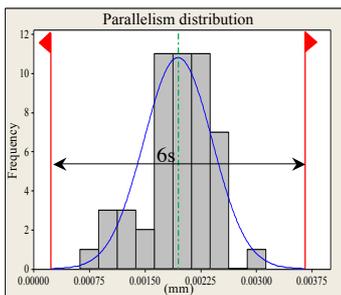

**Figure 13: Parallelism distribution**

| Name | Parallelism |
|---|---|
| Mean $\bar{x}$ (mm) | $1.944\times10^{-3}$ |
| Standard deviation $s$ (mm) | $0.461\times10^{-3}$ |

**Table 6: Parallellism value**

The results of the flatness analysis affirm once again that the quality of the machined planes 2 is better than the other one. It can be seen that the standard deviation of the parallelism between the two planes is not far from the standard deviation of the flatness of the planes 2; it means that the parallelism errors are insignificant.

### 5.2.3 – Process capability analysis

Final products have to satisfy customer requirements, which are defined using a specification limit (T). A comparison of the specification limit with total process variation (6s) will be used in order to estimate whether a capable process meets the above requirements. This comparison gives an obvious process capability index; know as the $C_p$ of the process [J]. In other words, the capability index shows how well the data fits in the specification limit. It is easy to see that "the higher the value, the better the fit". Normally, the $C_p$ values of 1.33 or greater are recommended.

Here, a batch of 50 parts is analyzed, therefore the $C_p$ is a sample estimator and is denoted $\hat{C}_p$.

$$\hat{C}_p = \frac{T}{6s} \qquad (6)$$

In order to use this equation (6), the distribution of a sample estimator is well know under normality.

In our case, the requirements can be flatness and parallelism tolerance of machined planes as follows (Figure 14):

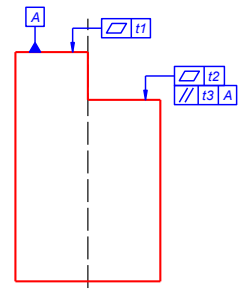

**Figure 14: Flatness and parallelism tolerances**

If values of the tolerances are known, it is easy to determine the process capability indices. For instance, if the values of the flatness and parallelism tolerances of the two machined planes are given as in the following table, the values of $\hat{C}_p$ can be obtained.

| Name | Flatness tolerance | | Parallelism tolerance ($t_3$) |
|---|---|---|---|
| | $t_1$ | $t_2$ | |
| T (mm) | 0.05 | 0.04 | 0.04 |
| $\hat{C}_p$ | 18.19 | 14.44 | 14.46 |

**Table 7 : Process capability indices**

As it is aforementioned, the values of $\hat{C}_p$ obtained in the above example are greater than 1.33 so that the capable process satisfies the customer requirements.

Conversely, if the process capability indices are known the minimum of tolerances for manufacturing can be calculated. An example is now considered to guarantee the $C_p$=2 for flatness of machined planes 1 so that the flatness tolerance is obtained as follows:

$$t_1 = C_p \cdot 6s = 2\times6\times4.58\times10^{-4} \qquad (7)$$
$$\Rightarrow t_1 = 5.496\times10^{-3} \ (mm)$$

## 6-Conclusions

In conclusion, this study presents the two approaches that can be used to complete the analyses of machining defects.

The first one is used to determine the dispersions of displacements (translations, rotations) of the machined planes. The results have shown that tool path is one of the factors that can influence defects of milled surfaces. Besides, the results ($s^2_{R_X}, s^2_{R_Y}, s^2_{T_Z}$) can be used to simulate and validate machining processes, which will be the point of further investigations, or predict manufacturing defects.

The second one is applied to analyze types of form errors (comber, twist, and undulation), the flatness and the orientation defects (parallelism). In this approach, the results of the analyses (Figure 8 and 10) have shown that the form errors are more important than the displacement defects. Thus, causes that influence form errors are needed to be considered in this case. Among of the causes should be single out the types of tool paths, rigidity of milling tools or geometrical defaults of machines. In addition, the restricted number of points used for the form parameterization method must be highlight (only 10 points are obtained by measurements on the milling machine during machining). Even though, this number appears to be low, the modes of the form defects are almost obtained. These results can be used, not only to estimate qualities of the machined surfaces, but also to aid manufacturers in improving product quality. For example, principal causes of the form errors are rigidity of mill tools, spindle of machine, and then machining programs (tool paths, feed rate, etc.), machine operations, etc.

Finally, the process capability indices ($C_p$, $C_{pk}$) can also be obtained in order to verify custom requirements. Conversely, if the process capability indices are known the minimum of tolerances can be assessed.

The following table shows functions of the two approaches in this study:

| Name | Types of tolerances | | | | | Process capability indices | |
|---|---|---|---|---|---|---|---|
| | Small displacements | | Form | Orientation | Location | | |
| | Translation | Rotations | Flatness | Parallelism | Position | $C_p$ | $C_{pk}$ |
| SDT | ✗ | ✗ | | ✓ | ✓ | ✗ | ✓ |
| F.P. | | | ✗ | ✗ | ✓ | ✗ | ✓ |

Table 8 : Functions of the two approaches

*Notations:*

| | |
|---|---|
| ✗ | The calculations were obtained. |
| ✓ | The calculations can be obtained. |
| SDT | The Small Displacement Torsor method |
| F.P. | The form parameterization method |
| $C_p$ | Capability index |
| $C_{pk}$ | The centering capability index |

## 7- References